\documentclass{article} 
\usepackage{iclr2021_conference,times}
\usepackage[square,numbers]{natbib}
\usepackage{hyperref}
\usepackage{url}
\usepackage{graphicx}
\usepackage{booktabs}
\usepackage{threeparttable}
\usepackage{amsmath}
\usepackage{amsfonts}
\usepackage{listings}
\usepackage{multirow}
\usepackage{threeparttable}

\title{Benchmarking Accuracy and Generalizability of Four Graph Neural Networks Using Large In Vitro ADME Datasets from Different Chemical Spaces}
\author{
Fabio Broccatelli \textsuperscript{1},
Richard Trager \textsuperscript{1},
Michael Reutlinger \textsuperscript{2},
George Karypis \textsuperscript{3, 4},
Mufei Li \textsuperscript{5}\\
\textsuperscript{1} Genentech, \textsuperscript{2} Roche Innovation Center Basel, \textsuperscript{3} AWS AI \\
\textsuperscript{4} University of Minnesota, \textsuperscript{5} AWS Shanghai AI Lab
}

\iclrfinalcopy
\begin{document}

\maketitle

\begin{abstract}
In this work, we benchmark a variety of single- and multi-task graph neural network (GNN) models against lower-bar and higher-bar traditional machine learning approaches employing human engineered molecular features. We consider four GNN variants -- Graph Convolutional Network (GCN), Graph Attention Network (GAT), Message Passing Neural Network (MPNN), and Attentive Fingerprint (AttentiveFP). So far deep learning models have been primarily benchmarked using lower-bar traditional models solely based on fingerprints, while more realistic benchmarks employing fingerprints, whole-molecule descriptors and predictions from other related endpoints (e.g., LogD7.4) appear to be scarce for industrial ADME datasets. In addition to time-split test sets based on Genentech data, this study benefits from the availability of measurements from an external chemical space (Roche data). We identify GAT as a promising approach to implementing deep learning models. While all GNN models significantly outperform lower-bar benchmark traditional models solely based on fingerprints, only GATs seem to offer a small but consistent improvement over higher-bar benchmark traditional models. Finally, the accuracy of in vitro assays from different laboratories predicting the same experimental endpoints appears to be comparable with the accuracy of GAT single-task models, suggesting that most of the observed error from the models is a function of the experimental error propagation.
\end{abstract}

\section{Introduction}

Absorption, distribution, metabolism and elimination (ADME) of new chemical entities were recognized in the 90' as the leading cause for clinical attribution of drug candidates~\cite{Kola2008}. This challenge was significantly reduced thanks to the advancements in the area of in vitro ADME modeling, which allows predicting the pharmacokinetic profile of novel compounds based on in vitro biological assays. These advancements were in turn largely driven by the availability of reliable in vitro reagents and advancement in LC-MS instrumentation. At Genentech every new project compound is tested in three ``first tier'' ADME assays (kinetic solubility, metabolic stability in microsomes, lipophilicity). A fraction of the most promising compounds is additionally tested in more biologically relevant ``second tier'' assays (e.g., metabolic stability in hepatocytes and permeability) and ``third tier'' assays that are less amenable for high-throughput screening (e.g., long term culture stability assays).

Quantitative structure activity relationship (QSAR) models are routinely utilized across industry to prioritize compounds to be synthesized or tested~\cite{ADME-PK}. These machine learning (ML) models are employed as ``zero tier'' assays to prioritize ideas based on the chemical structures. QSAR models allow exploring and de-risking a large portion of neighboring chemical space, which is beneficial due to the high cost of chemical synthesis. Before the use of deep learning (DL) was popularized in chemistry, the choice of the human engineered molecular features (or descriptors) was typically found to be the most important consideration to achieve accurate ML ADME models~\cite{ADME-PK}. This was particularly true for small datasets for which human engineered descriptors bespoke to the predicted property~\cite{Patrizia}. With the availability of larger datasets and GPU based technologies, reliance on ``chemically meaningful'' descriptors has decreased and machine engineered features produced by graph neural networks (GNNs) have gained popularity due to the ability to develop data driven representations leading to apparent improvement in accuracy and scalability~\cite{PotentialNet, Weave, Xiong19}.

The strategy behind in vitro ADME data gathering across different companies is guided by trying to maximize assay biological relevance versus assay cost. Frequently the more biologically relevant assays require more resources and are less amenable for high-throughput experiments. Consequently, datasets for the more biologically relevant assays tend to be smaller than datasets for workhorse ``first tier'' assays, which often aim to predict related endpoints (e.g. metabolic turnover). This sequential approach to compound characterization offers opportunities for transfer learning. Model architectures capable of leveraging information contained in higher volume assays in models for lower volume assays are desirable due to the potential increase in accuracy and ability to generalize.

Several benchmark studies have explored the usefulness of employing neural networks and graph convolution in QSAR models~\cite{PotentialNet, Weave, Kearnes17, Ramsundar15, Ramsundar17, Xiong19}. These studies demonstrated that single- and multi-task DL models substantially outperform single-task traditional QSAR models employing molecular fingerprints over a variety of tasks. Internally at Genentech we have observed that models exclusively based on fingerprints tend to underperform, and we have historically adopted a tiered ``dependency based'' model architecture. In this architecture, models are trained sequentially, and predictions of some models are used as molecular features in the following tier of models (e.g. calculated lipophilicity is utilized as a feature when predicting solubility). While previously published works applying DL to ADME datasets are encouraging, additional studies which better reflect industry practices may be required to adequately benchmark the performance of such models.

In this study, we evaluate the utility of GNNs to predict ADME properties solely based on molecular structures. We evaluate these models on four internal datasets, which are selected based on their relatively large size, importance in the decision-making across all Genentech projects, and assay interdependency (to better address the potential for transfer learning). Two of the four datasets are accompanied with an additional subset from the external chemical space (Roche data), enabling us to evaluate the ability of the models to generalize outside of the neighboring chemical space. In this study, four GNN variants are evaluated in single- and multi-task mode~\cite{Ramsundar17}. We compare these results against both lower-bar and higher-bar traditional ML approaches. Both of these benchmarks employ single-task dependency free extremely randomized trees models. The lower-bar models are only based on fingerprints, whereas the higher-bar models consider additional RDKit descriptors. The selection of extremely randomized trees as non-DL benchmark model, as opposed to more traditional random forest models, emerges from extensive in house evaluations (not discussed in this work).

\section{Experimental Section}

\subsection{Datasets}

Four internal datasets are used to train and test models: lipophilicity (LogD7.4), clearance intrinsic in human liver microsomes (HLM $\textsc{CL}_{int}$), clearance intrinsic in hepatocytes (HH $\textsc{CL}_{int}$), kinetic solubility phosphate buffer (KinSol)~\cite{Aliagas15, Halladay07, Lin13, Lin16}. The internal datasets are split into training and test sets based on the date of testing (time-split). The test sets consist of the compounds tested across a time period of approximately 3 months with the chronological cut-off applied~\cite{Sheridan13}. Genentech and Roche are part of the same organization, although they have historically operated independently. Recently large datasets of ADME data (LogD7.4 and HLM $\textsc{CL}_{int}$) have been shared. This creates the opportunity to test internal Genentech models on a large portion of unexplored chemical space, which in turn allows investigating the ability of the different models to generalize.

All the endpoints are transformed in logarithmic scale. Due to underlying relationships between different assays, some models may benefit from incorporating the predictions of other models into the input features during training and inference. For example, LogD7.4 is an important component of water solubility and is a driver for metabolic clearance in both HLM and HH~\cite{Broccatelli18}. Similarly, HLM contains most of the drug metabolizing enzymes present in HH. These relationships, defined as ``dependencies'', are human engineered (established based on the biological understanding of the mechanistic processes measured by the assay) and have been validated by years of experience in model deployment at Genentech~\cite{Aliagas15}. Table~\ref{tbl:dataset} describes the different datasets characterized in this study.

\begin{table}[t]
\begin{center}
\footnotesize
\caption{Datasets Used
\label{tbl:dataset}}
\begin{threeparttable}
\begin{tabular}{lccccc}
\toprule
Assay & Abbreviation & \# Train Set & \# Temporal Test Set & \# Roche Set & Dependencies \\ 
\midrule
LogD7.4 & LogD7.4 & 111,144 & 3,351 & 58,820 & \\
\midrule
Kinetic Solubility\\ (uM) at pH 7.4 & KinSol & 99,999 & 2,915 & & LogD7.4 \\
\midrule
Metabolic Stability\\ in human liver\\ microsome (CLint) & HLM $\textsc{CL}_{int}$ & 76,556 & 2,984 & 267,120 & LogD7.4 \\
\midrule
Metabolic Stability\\ in isolated human\\ hepatocytes (CLint) & HH $\textsc{CL}_{int}$ & 18,862 & 476 & & LogD7.4, HLM $\textsc{CL}_{int}$\\
\bottomrule
\end{tabular}
\begin{tablenotes}[flushleft]\footnotesize
\item \# set is the number of datapoints in the set.
\end{tablenotes}
\end{threeparttable}
\normalsize
\end{center}
\end{table}

In order to quantify the chemical diversity between the temporal test set and the Roche set as compared against the training set, we calculated the maximum Tanimoto similarity based on chemical fingerprints using a subset of the compounds ($20\%$ random selection stratified by the 3 different sets). This analysis was performed for both LogD7.4 and HLM $\textsc{CL}_{int}$. The average maximum similarity against the training set was $0.64$ (standard deviation equals to 0.14) for the temporal test set and $0.42$ (standard deviation equals to 0.06) for the Roche set.

\subsection{Methods}

\paragraph{Graph Construction and Featurization for GNNs.} We construct molecular graphs for GNNs, with atoms being nodes and bonds being edges. RDKit~\cite{rdkit} is used to generate the set of atom and bond descriptors in Table~\ref{tbl:descriptor}, which were used as input node and edge features for GNNs.

\begin{table}[t]
\begin{center}
\footnotesize
\caption{Atom and Bond Descriptors for Featurization
\label{tbl:descriptor}}
\begin{threeparttable}
\begin{tabular}{lc}
\toprule
Descriptor & Value Range \\ 
\midrule
Atom degree & 1, 2, 3, 4, 6 \\
Atom type & B, Br, C, Cl, F, H, I, N, O, P, S, Se, Si \\
Atom chiral tag & CCW, CW, unspecified, other \\
Atom formal charge & -1, 0, 1 \\
Atom hybridization & S, SP, SP2, SP3, SP3D2 \\
Atom implicit valence & 0, 1, 2, 3 \\
Aromatic atom & True, False \\
Atom mass & real numbers \\
\midrule
Bond type & aromatic, single, double, triple \\
Conjugated bond & True, False \\
Bond is in ring & True, False \\
Bond stereo configuration & OE, NONE, OZ, CIS, TRANS, ANY \\
\bottomrule
\end{tabular}
\end{threeparttable}
\normalsize
\end{center}
\end{table}

\paragraph{GNNs.} We develop the models with Deep Graph Library (DGL)~\cite{wang2020dgl} and DGL-LifeSci~\cite{dgllife}. We consider four GNN variants, including Graph Convolutional Network (GCN)~\cite{Kipf17}, Graph Attention Network (GAT)~\cite{Velivckovic18}, Message Passing Neural Network (MPNN)~\cite{pmlr-v70-gilmer17a}, and Attentive Fingerprint (AttentiveFP)~\cite{Xiong19}. While GCN and GAT were originally developed for node classification, DGL-LifeSci extends them for graph property prediction.

\paragraph{Multi-task DL Architectures.} We considered two architectures for multi-task learning, inspired by Ramsundar et al.~\cite{Ramsundar17}. The \textbf{parallel architecture} is the same as the architecture for single-task prediction, except that the multilayer perceptron (MLP) has multiple outputs, one per task. The \textbf{bypass} architecture employs $T+1$ GNNs for prediction on $T$ tasks, one shared by all tasks and one additional GNN for each task. For prediction on task $t$, the input is passed to both the shared GNN and the task-specific GNN. Output representations of the two GNNs are concatenated and then passed on to the prediction pipeline as in the parallel architecture, except that there is no parameter sharing across tasks. Different from the classical parallel architecture, the bypass architecture allows developing both task-shared chemical representations and task-specific ones.

\paragraph{Training of DL Models.} Smooth L1 loss was used as the loss function. Adam with weight decay was used for optimization. Early stopping was performed based on the squared Pearson correlation coefficient averaged over tasks for the internal validation set. The hyperparameters were chosen based on the best internal validation performance observed. We used Bayesian optimization with hyperopt~\cite{Bergstra12} for hyperparameter search, with the number of trials being 20.

\subsection{Baseline Models and Assays}

Baseline models were created using ExtraTreesRegressor (XRT)\footnote{\href{https://scikit-learn.org/stable/modules/generated/sklearn.ensemble.ExtraTreesRegressor.html}{https://scikit-learn.org/stable/modules/generated/sklearn.ensemble.ExtraTreesRegressor.html}} from scikit-learn~\cite{scikit-learn}. The number of trees was 100, the minimum number of samples required to form a leaf was 1, and the maximum tree depth was None. The hyperparameters were chose based on extensive internal evaluations not described in this work. 

We trained two types of baseline models. The \textbf{lower-bar} variant (XRT\rule{0.15cm}{0.15mm} FP) was only based on RDKit Morgan fingerprints (4096 bits, radius 2). The \textbf{higher-bar} variant (XRT\rule{0.15cm}{0.15mm} All) was based on both Morgan fingerprints and additional RDKit descriptors. The higher-bar models were sequentially trained with the dependencies described in Table~\ref{tbl:dataset}. The additional descriptors were Chi0, Chi0n, Chi0v, Chi1, Chi1n, Chi1v, Chi2n, Chi2v, Chi3n, Chi3v, Chi4n, Chi4v, EState\rule{0.15cm}{0.15mm}VSA1, EState\rule{0.15cm}{0.15mm} VSA2, EState\rule{0.15cm}{0.15mm} VSA3, EState\rule{0.15cm}{0.15mm} VSA4, EState\rule{0.15cm}{0.15mm} VSA5, EState\rule{0.15cm}{0.15mm} VSA6, EState\rule{0.15cm}{0.15mm} VSA7, EState\rule{0.15cm}{0.15mm} VSA8, EState\rule{0.15cm}{0.15mm} VSA9, EState\rule{0.15cm}{0.15mm} VSA10, EState\rule{0.15cm}{0.15mm} VSA11, FractionCSP3, HallKierAlpha, HeavyAtomCount, Kappa1, Kappa2, Kappa3, LabuteASA, MolLogP, MolMR, MolWt, NHOHCount, NOCount, NumAliphaticCarbocycles, NumAliphaticHeterocycles, NumAliphaticRings, NumAromaticCarbocycles, NumAromaticHeterocycles, NumAromaticRings, NumHAcceptors, NumHDonors, NumHeteroatoms, NumRotatableBonds, NumSaturatedCarbocycles, NumSaturatedHeterocycles, NumSaturatedRings, PEOE\rule{0.15cm}{0.15mm} VSA1, PEOE\rule{0.15cm}{0.15mm} VSA2, PEOE\rule{0.15cm}{0.15mm} VSA3, PEOE\rule{0.15cm}{0.15mm} VSA4, PEOE\rule{0.15cm}{0.15mm} VSA5, PEOE\rule{0.15cm}{0.15mm} VSA6, PEOE\rule{0.15cm}{0.15mm} VSA7, PEOE\rule{0.15cm}{0.15mm} VSA8, PEOE\rule{0.15cm}{0.15mm} VSA9, PEOE\rule{0.15cm}{0.15mm} VSA10, PEOE\rule{0.15cm}{0.15mm} VSA11, PEOE\rule{0.15cm}{0.15mm} VSA12, PEOE\rule{0.15cm}{0.15mm} VSA13, PEOE\rule{0.15cm}{0.15mm} VSA14, RingCount, SMR\rule{0.15cm}{0.15mm} VSA1, SMR\rule{0.15cm}{0.15mm} VSA2, SMR\rule{0.15cm}{0.15mm} VSA3, SMR\rule{0.15cm}{0.15mm} VSA4, SMR\rule{0.15cm}{0.15mm} VSA5, SMR\rule{0.15cm}{0.15mm} VSA6, SMR\rule{0.15cm}{0.15mm} VSA7, SMR\rule{0.15cm}{0.15mm} VSA8, SMR\rule{0.15cm}{0.15mm} VSA9, SMR\rule{0.15cm}{0.15mm} VSA10, SlogP\rule{0.15cm}{0.15mm} VSA1, SlogP\rule{0.15cm}{0.15mm} VSA2, SlogP\rule{0.15cm}{0.15mm} VSA3, SlogP\rule{0.15cm}{0.15mm} VSA4, SlogP\rule{0.15cm}{0.15mm} VSA5, SlogP\rule{0.15cm}{0.15mm} VSA6, SlogP\rule{0.15cm}{0.15mm} VSA7, SlogP\rule{0.15cm}{0.15mm} VSA8, SlogP\rule{0.15cm}{0.15mm} VSA9, SlogP\rule{0.15cm}{0.15mm} VSA10, SlogP\rule{0.15cm}{0.15mm} VSA11, SlogP\rule{0.15cm}{0.15mm} VSA12, TPSA, VSA\rule{0.15cm}{0.15mm} EState1, VSA\rule{0.15cm}{0.15mm} EState2, VSA\rule{0.15cm}{0.15mm} EState3, VSA\rule{0.15cm}{0.15mm} EState4, VSA\rule{0.15cm}{0.15mm} EState5, VSA\rule{0.15cm}{0.15mm} EState6, VSA\rule{0.15cm}{0.15mm} EState7, VSA\rule{0.15cm}{0.15mm} EState8, VSA\rule{0.15cm}{0.15mm} EState9, VSA\rule{0.15cm}{0.15mm} EState10.

Additional available RDKit descriptors were excluded from the selection due to the observed potential to generate numerical and non-numerical values that were problematic to handle within the modeling framework.

For a limited number of compounds in the LogD7.4 (139) and HLM $\textsc{CL}_{int}$ (45) datasets, experimental data was available in both the Genentech assay and the Roche assay. These sets were utilized to quantify the extent of variability deriving from using related but distinct assay protocols.

\section{Results and Discussion}

The performance of the different models across the temporal test set and the Roche set was evaluated according to three statistical metrics -- average error (AvgError), squared Pearson correlation coefficient ($r^2$), and the fraction of predictions within 1 log unit (within1Log).

In order to reduce the combinatorial explosion resulting from four distinct GNN variants, two multi-task model architectures, and four datasets, single-task models were approached initially as in Table~\ref{tbl:single}. GATs outperform the other models and are therefore utilized in multi-task models to evaluate the ability of inter-task transfer learning.

\begin{table}[t]
\begin{center}
\footnotesize
\caption{Performance of Single-Task Models
\label{tbl:single}}
\begin{threeparttable}
\begin{tabular}{lcccc}
\toprule
Model & Dataset & AvgError & $r^2$ & within1Log \\ 
\midrule
GCN & LogD7.4 Temporal & \textbf{0.58} & \textbf{0.65} & \textbf{0.87} \\
GAT & LogD7.4 Temporal & 0.66 & 0.63 & 0.82 \\
MPNN & LogD7.4 Temporal & 0.61 & 0.62 & 0.84 \\
AttentiveFP & LogD7.4 Temporal & \textbf{0.58} & 0.56 & \textbf{0.87} \\
\midrule
GCN & LogD7.4 Roche & 0.62 & \textbf{0.65} & 0.81 \\
GAT & LogD7.4 Roche & \textbf{0.60} & 0.63 & \textbf{0.83} \\
MPNN & LogD7.4 Roche & 0.67 & 0.60 & 0.79 \\
AttentiveFP & LogD7.4 Roche & 0.68 & 0.61 & 0.77 \\
\midrule
GCN & HLM $\textsc{CL}_{int}$ Temporal & 0.39 & 0.44 & 0.94 \\
GAT & HLM $\textsc{CL}_{int}$ Temporal & \textbf{0.38} & 0.48 & \textbf{0.95} \\
MPNN & HLM $\textsc{CL}_{int}$ Temporal & 0.40 & 0.41 & 0.93 \\
AttentiveFP & HLM $\textsc{CL}_{int}$ Temporal & 0.47 & \textbf{0.53} & 0.89 \\
\midrule
GCN & HLM $\textsc{CL}_{int}$ Roche & 0.45 & 0.19 & 0.93 \\
GAT & HLM $\textsc{CL}_{int}$ Roche & \textbf{0.42} & \textbf{0.24} & \textbf{0.95} \\
MPNN & HLM $\textsc{CL}_{int}$ Roche & 0.49 & 0.16 & 0.91 \\
AttentiveFP & HLM $\textsc{CL}_{int}$ Roche & 0.46 & 0.19 & 0.92 \\
\midrule
GCN & KinSol Temporal & 0.56 & 0.22 & 0.86 \\
GAT & KinSol Temporal & \textbf{0.49} & \textbf{0.26} & \textbf{0.89} \\
MPNN & KinSol Temporal & 0.53 & 0.24 & 0.86 \\
AttentiveFP & KinSol Temporal & 0.56 & 0.25 & 0.85 \\
\midrule
GCN & HH $\textsc{CL}_{int}$ Temporal & 0.42 & 0.39 & 0.93 \\
GAT & HH $\textsc{CL}_{int}$ Temporal & 0.39 & \textbf{0.45} & \textbf{0.98} \\
MPNN & HH $\textsc{CL}_{int}$ Temporal & \textbf{0.37} & 0.41 & 0.96 \\
AttentiveFP & HH $\textsc{CL}_{int}$ Temporal & 0.39 & 0.37 & 0.95 \\
\midrule
GCN & Averaged & 0.50 & 0.42 & 0.89 \\
GAT & Averaged & \textbf{0.49} & \textbf{0.45} & \textbf{0.90} \\
MPNN & Averaged & 0.51 & 0.41 & 0.88 \\
AttentiveFP & Averaged & 0.52 & 0.42 & 0.88 \\
\midrule
GCN & Averaged, excluding LogD7.4 & 0.46 & 0.31 & 0.92 \\
GAT & Averaged, excluding LogD7.4 & \textbf{0.42} & \textbf{0.36} & \textbf{0.94} \\
MPNN & Averaged, excluding LogD7.4 & 0.45 & 0.30 & 0.92 \\
AttentiveFP & Averaged, excluding LogD7.4 & 0.47 & 0.34 & 0.90 \\
\bottomrule
\end{tabular}
\end{threeparttable}
\normalsize
\end{center}
\end{table}

\begin{figure}[tb!]
  \centering
  \includegraphics[width=\textwidth]{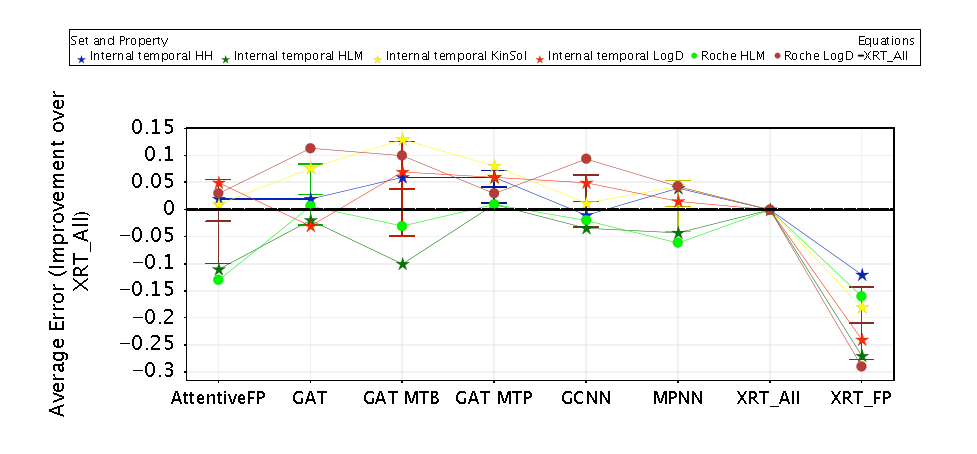}
  \caption{Average error of the different models analyzed across multiple datasets as compared with respect to XRT\rule{0.15cm}{0.15mm} All. Stars represent temporal test sets, while circles represent external Roche sets. Green: HLM $\textsc{CL}_{int}$, Red: LogD7.4, Blue: HH $\textsc{CL}_{int}$, Yellow: KinSol}
  \label{fig:difference_All}
\end{figure}

Table~\ref{tbl:gnn_baseline} characterizes the performance of all GAT and XRT benchmark models analyzed in this study across 6 test sets (4 temporal, 2 external). GAT ST, GAT MTP and GAT MTB are respectively single-task GAT, multi-task GAT with parallel architecture, and multi-task GAT with bypass architecture. While some differences can be observed from dataset to dataset, the conclusions are fairly consistent overall. Fingerprint-only models perform significantly worse than all the other methodologies ($p < 0.05$ for all the metrics) and thus are not an adequate benchmark for DL models. The average error of fingerprint-only models across all datasets equals $0.73$, which is substantially higher than that of the other traditional and DL models (average error between $0.48$ and $0.52$). The gap in performance between the best overall methodologies (GAT MTB and GAT MTP) and traditional models using additional descriptors and dependencies (XRT\rule{0.15cm}{0.15mm} All) is significantly narrower (average error $0.48$ vs $0.52$). By analyzing the performance difference between all methods with respect to XRT\rule{0.15cm}{0.15mm} All as in Figure~\ref{fig:difference_All}, GAT MTP is the only model for which a statistically significant improvement is observed across all metrics. GAT MTB appears to be more variable and requires more time for training and hyperparameter tuning. As a result, this architecture is less reliable. While GAT ST models are on average superior to XRT\rule{0.15cm}{0.15mm} All models, the improvement appears incremental and is not sufficient to achieve statistical significance. Interestingly, these conclusions do not hold when the two external datasets (Roche data) are analyzed. These datasets are a higher bar because a higher accuracy for them implies a stronger ability to generalize. In addition, their size is considerably larger than the temporal test sets. GAT ST models appear to display the best compromise between local accuracy and generalizability, and are therefore used as the preferred method for subsequent analysis in this work. More comprehensive analysis including additional endpoints and external datasets might be required to cement these conclusions.

\begin{table}[t]
\begin{center}
\footnotesize
\caption{Performance of All GAT and XRT Models
\label{tbl:gnn_baseline}}
\begin{threeparttable}
\begin{tabular}{lcccc}
\toprule
Model & Dataset & AvgError & $r^2$ & within1Log \\ 
\midrule
GAT ST & LogD7.4 Temporal & 0.66 & \textbf{0.63} & 0.82 \\
GAT MTP & LogD7.4 Temporal & \textbf{0.56} & 0.57 & \textbf{0.89} \\
GAT MTB & LogD7.4 Temporal & \textbf{0.56} & 0.56 & 0.88 \\
XRT\rule{0.15cm}{0.15mm} FP & LogD7.4 Temporal & 0.87 & 0.25 & 0.68 \\
XRT\rule{0.15cm}{0.15mm} All & LogD7.4 Temporal & 0.63 & 0.55 & 0.83 \\
\midrule
GAT ST & LogD7.4 Roche & \textbf{0.60} & \textbf{0.63} & \textbf{0.83} \\
GAT MTP & LogD7.4 Roche & 0.68 & \textbf{0.63} & 0.77 \\
GAT MTB & LogD7.4 Roche & 0.61 & \textbf{0.63} & 0.81 \\
XRT\rule{0.15cm}{0.15mm} FP & LogD7.4 Roche & 1.00 & 0.14 & 0.52 \\
XRT\rule{0.15cm}{0.15mm} All & LogD7.4 Roche & 0.71 & 0.50 & 0.76 \\
\midrule
GAT ST & HLM $\textsc{CL}_{int}$ Temporal & 0.38 & 0.48 & 0.95 \\
GAT MTP & HLM $\textsc{CL}_{int}$ Temporal & \textbf{0.35} & 0.48 & \textbf{0.96} \\
GAT MTB & HLM $\textsc{CL}_{int}$ Temporal & 0.46 & \textbf{0.55} & 0.89 \\
XRT\rule{0.15cm}{0.15mm} FP & HLM $\textsc{CL}_{int}$ Temporal & 0.63 & 0.21 & 0.80 \\
XRT\rule{0.15cm}{0.15mm} All & HLM $\textsc{CL}_{int}$ Temporal & 0.36 & 0.45 & 0.95 \\
\midrule
GAT ST & HLM $\textsc{CL}_{int}$ Roche & \textbf{0.42} & \textbf{0.24} & 0.95 \\
GAT MTP & HLM $\textsc{CL}_{int}$ Roche & \textbf{0.42} & \textbf{0.24} & \textbf{0.96} \\
GAT MTB & HLM $\textsc{CL}_{int}$ Roche & 0.46 & 0.23 & 0.94 \\
XRT\rule{0.15cm}{0.15mm} FP & HLM $\textsc{CL}_{int}$ Roche & 0.59 & 0.03 & 0.82 \\
XRT\rule{0.15cm}{0.15mm} All & HLM $\textsc{CL}_{int}$ Roche & 0.43 & 0.17 & \textbf{0.96} \\
\midrule
GAT ST & KinSol Temporal & 0.49 & 0.26 & 0.89 \\
GAT MTP & KinSol Temporal & 0.49 & 0.27 & 0.91 \\
GAT MTB & KinSol Temporal & 0.44 & \textbf{0.30} & 0.93 \\
XRT\rule{0.15cm}{0.15mm} FP & KinSol Temporal & 0.75 & 0.02 & 0.73 \\
XRT\rule{0.15cm}{0.15mm} All & KinSol Temporal & \textbf{0.43} & 0.17 & \textbf{0.96} \\
\midrule
GAT ST & HH $\textsc{CL}_{int}$ Temporal & 0.39 & 0.45 & \textbf{0.98} \\
GAT MTP & HH $\textsc{CL}_{int}$ Temporal & \textbf{0.35} & 0.50 & \textbf{0.98} \\
GAT MTB & HH $\textsc{CL}_{int}$ Temporal & \textbf{0.35} & \textbf{0.54} & 0.95 \\
XRT\rule{0.15cm}{0.15mm} FP & HH $\textsc{CL}_{int}$ Temporal & 0.53 & 0.25 & 0.85 \\
XRT\rule{0.15cm}{0.15mm} All & HH $\textsc{CL}_{int}$ Temporal & 0.41 & 0.47 & 0.93 \\
\midrule
GAT ST & Averaged & 0.49 & 0.45 & 0.90 \\
GAT MTP & Averaged & \textbf{0.48} & 0.45 & \textbf{0.91} \\
GAT MTB & Averaged & \textbf{0.48} & \textbf{0.47} & 0.90 \\
XRT\rule{0.15cm}{0.15mm} FP & Averaged & 0.73 & 0.15 & 0.73 \\
XRT\rule{0.15cm}{0.15mm} All & Averaged & 0.52 & 0.40 & 0.86 \\
\midrule
GAT ST & Averaged, excluding LogD7.4 & 0.42 & 0.36 & 0.94 \\
GAT MTP & Averaged, excluding LogD7.4 & \textbf{0.40} & 0.37 & \textbf{0.95} \\
GAT MTB & Averaged, excluding LogD7.4 & 0.43 & \textbf{0.40} & 0.93 \\
XRT\rule{0.15cm}{0.15mm} FP & Averaged, excluding LogD7.4 & 0.63 & 0.13 & 0.80 \\
XRT\rule{0.15cm}{0.15mm} All & Averaged, excluding LogD7.4 & 0.44 & 0.33 & 0.93 \\
\bottomrule
\end{tabular}
\end{threeparttable}
\normalsize
\end{center}
\end{table}

The two datasets for LogD7.4 allow an interesting case study. Lipophilicity is an important component of, and is not influenced by, all the other properties modeled in this study. Consequently, the benefits of multi-task learning as a means to enabling transfer learning should theoretically be less apparent when modeling this endpoint. Interestingly, single-task GAT performed worse than the multi-task GAT and the high-bar XRT benchmark on the LogD7.4 temporal test set, while it was the best model on LogD7.4 Roche set, which is composed of over 59k compounds. The gap in average error observed on the temporal test set between single-task and multi-task GAT models was the broadest across all the datasets analyzed. While some degree of underfitting was observed in the local chemical space for single-task GAT, these shortcomings had little influence on the ability of the model to generalize to a different chemical space. Intrigued by this observation, we increased the number of trials in hyperparameter search from 20 to 40 for the special case of single-task GAT on LogD7.4. This improved the performance on both the temporal test set (average error from $0.66$ to $0.61$, same $r^2$, and fraction within 1 log unit from $0.82$ to $0.86$) and the Roche set (average error from $0.60$ to $0.58$, $r^2$ from $0.63$ to $0.66$, and fraction within 1 log unit from $0.83$ to $0.84$). For these two different chemical spaces, single-task GAT exhibited both similar accuracy and competitive performance, indicating a superior ability to generalize and a lower susceptibility to overfit to the local space. The performance of single-task GAT trained on Genentech data was additionally compared with the Genentech's assay ability of Genentech to predict data for compounds tested in the Roche assay as in Table~\ref{tbl:assay}. This comparison allows quantifying the composite experimental variability resulting from measuring compounds in a different laboratory with related but distinct protocol. Surprisingly the experimental assay and single-task GAT exhibited comparable accuracy for this dataset. 

The same analysis was performed for the HLM $\textsc{CL}_{int}$ dataset. In this case, the model exhibited a better average error (0.42 vs 0.44) but a significantly lower $r^2$ ($0.24$ vs $0.51$), reflecting the tendency of the model to populate the middle of the assay dynamic range and a lower ability to rank-order observations at the extremes. Nevertheless, the model is remarkably better than the assay in avoiding large prediction errors (fraction within 1 log unit $0.95$ vs $0.91$). Taken together, these observations point to that the in-silico model error is largely a product of the in vitro variability, in agreement with the work presented by Wenlock and Carlsson~\cite{Wenlock14}.

\begin{table}[t]
\begin{center}
\footnotesize
\caption{Performance of Genentech In-Silico Models and In Vitro Assays in Predicting Roche In Vitro Assay Data
\label{tbl:assay}}
\begin{threeparttable}
\begin{tabular}{lccccc}
\toprule
Property & Method & AvgError & $r^2$ & within1Log & N \\
\midrule
LogD7.4 & Single-Task GAT & 0.58 & \textbf{0.66} & 0.84 & 61,427 \\
LogD7.4 & Experimental & \textbf{0.55} & 0.63 & \textbf{0.86} & 139 \\
\midrule
HLM $\textsc{CL}_{int}$ & Single-Task GAT & \textbf{0.42} & 0.24 & \textbf{0.95} & 26,719 \\
HLM $\textsc{CL}_{int}$ & Experimental & 0.44 & \textbf{0.51} & 0.91 & 45 \\
\bottomrule
\end{tabular}
\end{threeparttable}
\normalsize
\end{center}
\end{table}

\section{Conclusions}

This study identifies multi-task GAT as a state of the art approach across most tasks. All the other single-task DL approaches (GCN, MPNN, AttentiveFP) evaluated did not appear to outperform traditional XRT models utilizing fingerprints, general descriptors and dependencies (e.g., using predictions from other models as descriptors). Similarly, single-task GAT models were found to be only incrementally better than traditional XRT models utilizing fingerprints, general descriptors and dependencies (XRT\rule{0.15cm}{0.15mm} All). In contrast, XRT models exclusively based on fingerprints performed significantly worse compared with all the other approaches across all tasks and datasets. DL models, GAT in particular, exhibited a higher accuracy compared with XRT models on the LogD7.4 datasets. For this task, the availability of the independent dataset composed of Roche data allowed differentiating single-task GAT in terms of the ability to generalize to an external chemical space. Strikingly, the Genentech assay for LogD7.4 was only slightly more accurate than the GAT model, suggesting that the model accuracy is approaching the limit imposed by the experimental assay error propagation. The other prediction tasks approached in this work (KinSol, HLM $\textsc{CL}_{int}$, HH $\textsc{CL}_{int}$) are influenced by LogD7.4, and were therefore utilized as benchmark for transfer learning. Surprisingly, multi-task models trained across different tasks in parallel outperform single-task GAT models by a minimal margin, which appears to be erased and reversed for the considerably larger and more diverse external Roche datasets. In aggregate, considering the limited improvement in performance for multi-task models in light of the high potential for transfer learning on these datasets, these findings suggest that transfer learning for GAT could be improved by alternative approaches. One potential approach could involve designing networks explicitly reproducing the hierarchy of inter-task relationships based on the knowledge of the assays. Furthermore, given the remarkable similarity in performance between assays and in silico models, it is possible that the lack of clear differentiation between most approaches explored in this study is a reflection of the proximity to the theoretical accuracy limit dictated by the assay error propagation.

This study highlights the importance of utilizing appropriate benchmarking in order to understand strengths and areas for opportunities for new technologies. In light of these findings, it is recommended that models solely based on fingerprints are not utilized as benchmarks for deep learning approaches in ADME. In addition, this study emphasizes the importance of using orthogonal approaches to model validation where possible. Despite our best efforts to validate models using a higher bar splitting paradigm (time-split), it is the availability of two large external datasets (Roche data for LogD7.4 and HLM $\textsc{CL}_{int}$) that ultimately allowed us to differentiate single-task GAT models for their ability to generalize (particularly for LogD7.4). From an algorithmic standpoint, these observations might be challenging to address since the opportunity to consistently test models over external chemical space may be missing for many experimental endpoints. From an operational standpoint, these findings reinforce the importance of in-depth outlier anlaysis, open communication with model users, and partnership with experimental scientists as tools to assess and improve model impact. Natural next steps for this work include the extension of the analysis to ADME assays that are less data rich and the experimentation of different approaches to transfer learning. The code and chosen hyperparameters are released on github \href{https://github.com/awslabs/dgl-lifesci/tree/master/examples/property_prediction/MTL}{here}.

\bibliography{main}

\begin{thebibliography}{24}
\providecommand{\natexlab}[1]{#1}
\providecommand{\url}[1]{\texttt{#1}}
\expandafter\ifx\csname urlstyle\endcsname\relax
  \providecommand{\doi}[1]{doi: #1}\else
  \providecommand{\doi}{doi: \begingroup \urlstyle{rm}\Url}\fi

\bibitem[Aliagas et~al.(2015)Aliagas, Gobbi, Heffron, Lee, Ortwine, Zak, and
  Khojasteh]{Aliagas15}
I.~Aliagas, A.~Gobbi, T.~Heffron, M.-L. Lee, D.~F. Ortwine, M.~Zak, and S.~C.
  Khojasteh.
\newblock A probabilistic method to report predictions from a human liver
  microsomes stability qsar model: a practical tool for drug discovery.
\newblock \emph{J. Comput.-Aided Mol. Des.}, 29\penalty0 (4):\penalty0
  327--338, 2015.

\bibitem[Bergstra et~al.(2012)Bergstra, Yamins, and Cox]{Bergstra12}
J.~Bergstra, D.~Yamins, and D.~D. Cox.
\newblock Making a science of model search: Hyperparameter optimization in
  hundreds of dimensions for vision architectures.
\newblock In \emph{Proceedings of the 30th International Conference on Machine
  Learning}, pages I--115--I--123, 2012.

\bibitem[Broccatelli et~al.(2018)Broccatelli, Aliagas, and
  Zheng]{Broccatelli18}
F.~Broccatelli, I.~Aliagas, and H.~Zheng.
\newblock Why decreasing lipophilicity alone is often not a reliable strategy
  for extending iv half-life.
\newblock \emph{ACS Med. Chem. Lett.}, 9\penalty0 (6):\penalty0 522--527, 2018.

\bibitem[Crivori et~al.(2000)Crivori, Cruciani, Carrupt, and Testa]{Patrizia}
P.~Crivori, G.~Cruciani, P.-A. Carrupt, and B.~Testa.
\newblock Predicting blood-brain barrier permeation from three-dimensional
  molecular structure.
\newblock \emph{J. Med. Chem.}, 43\penalty0 (11):\penalty0 2204--2216, 2000.

\bibitem[Feinberg et~al.(2019)Feinberg, Sheridan, Joshi, Pande, and
  Cheng]{PotentialNet}
E.~N. Feinberg, R.~Sheridan, E.~Joshi, V.~S. Pande, and A.~C. Cheng.
\newblock Step change improvement in admet prediction with potentialnet deep
  featurization.
\newblock \emph{arXiv preprint arXiv:1903.11789}, 2019.

\bibitem[Gilmer et~al.(2017)Gilmer, Schoenholz, Riley, Vinyals, and
  Dahl]{pmlr-v70-gilmer17a}
J.~Gilmer, S.~S. Schoenholz, P.~F. Riley, O.~Vinyals, and G.~E. Dahl.
\newblock Neural message passing for quantum chemistry.
\newblock In D.~Precup and Y.~W. Teh, editors, \emph{Proceedings of the 34th
  International Conference on Machine Learning}, volume~70 of \emph{Proceedings
  of Machine Learning Research}, pages 1263--1272. PMLR, 06--11 Aug 2017.

\bibitem[Halladay et~al.(2007)Halladay, Wong, Jaffer, Sinhababu, and
  Khojasteh-Bakht]{Halladay07}
J.~S. Halladay, S.~Wong, S.~M. Jaffer, A.~K. Sinhababu, and S.~C.
  Khojasteh-Bakht.
\newblock Metabolic stability screen for drug discovery using cassette analysis
  and column switching.
\newblock \emph{Drug Metab. Lett.}, 1\penalty0 (1):\penalty0 67--72, 2007.

\bibitem[Kearnes et~al.(2016)Kearnes, McCloskey, Berndl, Pande, and
  Riley]{Weave}
S.~Kearnes, K.~McCloskey, M.~Berndl, V.~Pande, and P.~Riley.
\newblock Molecular graph convolutions: moving beyond fingerprints.
\newblock \emph{J. Comput.-Aided Mol. Des.}, 30\penalty0 (8):\penalty0
  595--608, 2016.

\bibitem[Kearnes et~al.(2017)Kearnes, Goldman, and Pande]{Kearnes17}
S.~Kearnes, B.~Goldman, and V.~Pande.
\newblock Modeling industrial admet data with multitask networks.
\newblock \emph{arXiv preprint arXiv:1606.08793}, 2017.

\bibitem[Kipf and Welling(2017)]{Kipf17}
T.~N. Kipf and M.~Welling.
\newblock Semi-supervised classification with graph convolutional networks.
\newblock In \emph{International Conference on Learning Representations}, 2017.

\bibitem[Kola(2008)]{Kola2008}
I.~Kola.
\newblock The state of innovation in drug development.
\newblock \emph{Clin. Pharmacol. Ther.}, 83\penalty0 (2):\penalty0 227--230,
  2008.

\bibitem[Li et~al.(2021)Li, Zhou, Hu, Fan, Zhang, Gu, and Karypis]{dgllife}
M.~Li, J.~Zhou, J.~Hu, W.~Fan, Y.~Zhang, Y.~Gu, and G.~Karypis.
\newblock Dgl-lifesci: An open-source toolkit for deep learning on graphs in
  life science.
\newblock \emph{ACS Omega}, 2021.

\bibitem[Lin and Pease(2013)]{Lin13}
B.~Lin and J.~H. Pease.
\newblock A novel method for high throughput lipophilicity determination by
  microscale shake flask and liquid chromatography tandem mass spectrometry.
\newblock \emph{Comb. Chem. High T. Scr.}, 16\penalty0 (10):\penalty0 817--825,
  2013.

\bibitem[Lin and Pease(2016)]{Lin16}
B.~Lin and J.~H. Pease.
\newblock A high throughput solubility assay for drug discovery using
  microscale shake-flask and rapid uhplc-uv-clnd quantification.
\newblock \emph{J. Pharm. Biomed. Anal.}, 122:\penalty0 126--140, 2016.

\bibitem[Lombardo et~al.(2017)Lombardo, Desai, Arimoto, Desino, Fischer,
  Keefer, Petersson, Winiwarter, and Broccatelli]{ADME-PK}
F.~Lombardo, P.~V. Desai, R.~Arimoto, K.~E. Desino, H.~Fischer, C.~E. Keefer,
  C.~Petersson, S.~Winiwarter, and F.~Broccatelli.
\newblock In silico absorption, distribution, metabolism, excretion, and
  pharmacokinetics (adme-pk): Utility and best practices. an industry
  perspective from the international consortium for innovation through quality
  in pharmaceutical development.
\newblock \emph{J. Med. Chem.}, 60\penalty0 (22):\penalty0 9097--9113, 2017.

\bibitem[Pedregosa et~al.(2011)Pedregosa, Varoquaux, Gramfort, Michel, Thirion,
  Grisel, Blondel, Prettenhofer, Weiss, Dubourg, Vanderplas, Passos,
  Cournapeau, Brucher, Perrot, and Duchesnay]{scikit-learn}
F.~Pedregosa, G.~Varoquaux, A.~Gramfort, V.~Michel, B.~Thirion, O.~Grisel,
  M.~Blondel, P.~Prettenhofer, R.~Weiss, V.~Dubourg, J.~Vanderplas, A.~Passos,
  D.~Cournapeau, M.~Brucher, M.~Perrot, and E.~Duchesnay.
\newblock Scikit-learn: Machine learning in {P}ython.
\newblock \emph{Journal of Machine Learning Research}, 12:\penalty0 2825--2830,
  2011.

\bibitem[Ramsundar et~al.(2015)Ramsundar, Kearnes, Riley, Webster, Konerding,
  and Pande]{Ramsundar15}
B.~Ramsundar, S.~Kearnes, P.~Riley, D.~Webster, D.~Konerding, and V.~Pande.
\newblock Massively multitask networks for drug discovery.
\newblock \emph{arXiv preprint arXiv:1502.02072}, 2015.

\bibitem[Ramsundar et~al.(2017)Ramsundar, Liu, Wu, Verras, Tudor, Sheridan, and
  Pande]{Ramsundar17}
B.~Ramsundar, B.~Liu, Z.~Wu, A.~Verras, M.~Tudor, R.~P. Sheridan, and V.~Pande.
\newblock Is multitask deep learning practical for pharma?
\newblock \emph{J. Chem. Inf. Model.}, 57\penalty0 (8):\penalty0 2068--2076,
  2017.

\bibitem[RDKit()]{rdkit}
RDKit.
\newblock {RDK}it: Open-source cheminformatics.
\newblock \url{http://www.rdkit.org}.
\newblock [Online; accessed 7-Sep-2021].

\bibitem[Sheridan(2013)]{Sheridan13}
R.~P. Sheridan.
\newblock Time-split cross-validation as a method for estimating the goodness
  of prospective prediction.
\newblock \emph{J. Chem. Inf. Model.}, 53\penalty0 (4):\penalty0 783--790,
  2013.

\bibitem[Veli\v{c}kovi\'{c} et~al.(2018)Veli\v{c}kovi\'{c}, Cucurull, Casanova,
  Romero, Li\`{o}, and Bengio]{Velivckovic18}
P.~Veli\v{c}kovi\'{c}, G.~Cucurull, A.~Casanova, A.~Romero, P.~Li\`{o}, and
  Y.~Bengio.
\newblock Graph attention networks.
\newblock In \emph{International Conference on Learning Representations}, 2018.

\bibitem[Wang et~al.(2020)Wang, Zheng, Ye, Gan, Li, Song, Zhou, Ma, Yu, Gai,
  Xiao, He, Karypis, Li, and Zhang]{wang2020dgl}
M.~Wang, D.~Zheng, Z.~Ye, Q.~Gan, M.~Li, X.~Song, J.~Zhou, C.~Ma, L.~Yu,
  Y.~Gai, T.~Xiao, T.~He, G.~Karypis, J.~Li, and Z.~Zhang.
\newblock Deep graph library: A graph-centric, highly-performant package for
  graph neural networks.
\newblock \emph{arXiv preprint arXiv:1909.01315}, 2020.

\bibitem[Wenlock and Carlsson(2014)]{Wenlock14}
M.~C. Wenlock and L.~A. Carlsson.
\newblock How experimental errors influence drug metabolism and pharmacokinetic
  qsar/qspr models.
\newblock \emph{J. Chem. Inf. Model.}, 55\penalty0 (1):\penalty0 125--134,
  2014.

\bibitem[Xiong et~al.(2020)Xiong, Wang, Liu, Zhong, Wan, Li, Li, Luo, Chen,
  Jiang, and Zheng]{Xiong19}
Z.~Xiong, D.~Wang, X.~Liu, F.~Zhong, X.~Wan, X.~Li, Z.~Li, X.~Luo, K.~Chen,
  H.~Jiang, and M.~Zheng.
\newblock Pushing the boundaries of molecular representation for drug discovery
  with the graph attention mechanism.
\newblock \emph{J. Med. Chem.}, 63\penalty0 (16):\penalty0 8749--8760, 2020.

\end{thebibliography}

\end{document}